\documentclass[12pt]{iopart}

\def\th{\theta}
\def\ra{\rightarrow}

\newcommand{\be}{\begin{equation}}
\newcommand{\ee}{\end{equation}}
\newcommand{\bea}{\begin{eqnarray}}
\newcommand{\eea}{\end{eqnarray}}

\def\l{\lambda}

\def\none{${\cal N}=1$ }
\def\ntwo{${\cal N}=2$ }
\begin{document}
\begin{titlepage}

\hfill IC/2003/161
\par\hfill CPHT-PC-113-1203\\
\vskip .1in \hfill hep-th/0401045

\hfill

\vspace{20pt}

\title{PP-waves and softly broken ${\cal N}=1$ SYM}

\author{F. Bigazzi\dag\ and A. L. Cotrone\ddag}

\address{\dag\ The Abdus Salam International Centre for Theoretical Physics, Strada
Costiera, 11; I-34014 Trieste, Italy.}
\address{\ddag\ Centre de Physique Th\'eorique, \`Ecole Polytechnique, 48 Route de Saclay; F-91128 Palaiseau Cedex, France.\\
INFN, Piazza dei Caprettari, 70; I-00186  Roma, Italy.}

\begin{abstract}
We review the Penrose limit of the Type IIB dual of softly broken N=1 SYM in four dimensions obtained as a deformation of the Maldacena-N\`u\~nez background. We extract the string spectrum on the resulting pp-wave background and discuss some properties of the conjectured dual gauge theory hadrons, the so called ``Annulons''.
\end{abstract}

\eads{\mailto{bigazzif@ictp.trieste.it}, \mailto{Cotrone@cpht.polytechnique.fr}}

\end{titlepage}

\maketitle

\setcounter{page}{1}
\section{Motivations}

The predictive power of the string/gauge theory duality has long been limited mainly to the supergravity limit of the string theory, because the latter is generically very difficult to be solved on curved RR backgrounds. 
Apart from relegating the conformal field theories to the strong coupling regime of the planar limit, this limitation has forbidden the study of pure confining theories, which would need a full (in $\alpha'$) string theory description.
Recently, the situation has been improved by a simple proposal in the papers \cite{BMN,GKP}: apart from the rank $N$ of the gauge group and the 't Hooft coupling $\l$, the field theories have other quantities, such as the spin or symmetry charge quantum numbers, which allow for a ``perturbative expansion''.
In the regime where these quantities are very large, the dual string theory relevant objects are semiclassical solitons, thus quite simple to be studied.
In the mostly investigated case of the $AdS/CFT$ correspondence, for example, long spinning strings in $AdS$ describe large bare dimension and large spin operators, while point-like strings rotating on $S^5$ correspond to large bare dimension and large R-charge operators.
The latter case, for only one R-charge, is equivalent to the study of the string theory on the Penrose limit of $AdS_5 \times S^5$, which turns out to be solvable.

The natural question is then if we can extend these powerfull techniques to other, more interesting field theories.
In this paper, we will in fact review how we can extract non trivial data from the Penrose limit technique\footnote{The semiclassical solitonic approach for these backgrounds has been used in \cite{tsey}.} applied to backgrounds which are dual to confining and non supersymmetric theories (see \cite{noirev} for recent reviews on these dualities).
Thus, the relevant field theory objects under investigation are no more operators, but a particular type of {\it hadrons}.
These will turn out to be very heavy states with a very large symmetry charge, and, surprisingly, to have the physical shape of strings, with stringy excitations in the special directions \cite{pandostras,noi}.

The paper is organized as follows.
In section 2 we will introduce the relevant supergravity background, which is the softly broken version of the Maldacena-N\`u\~nez one\footnote{A very similar study can be performed in the softly broken Klebanov-Strassler case \cite{ks,ric}, but since the formulae are more complicate, we won't describe it here. The interested reader can look at the original papers \cite{pandostras,noi,sonne}.}, and we will perform the Penrose limit in the infrared region.
We will then solve the string theory on the resulting pp-wave, working out the spectrum, and we will discuss the stability of the model.
Then, in section \ref{data} we will study the main features of these hadrons, identifying the meaning of some stringy modes in the field theory, and we will end with some remarks in section \ref{conc}.

\section{String theory model}

\subsection{The supergravity background}
The softly broken version of the Maldacena-N\`u\~nez background \cite{mn,gubsermn,epz} has the form
\bea
\label{MNsol}
\fl ds^2_{str} =  e^{\Phi } \left[ dx_\mu dx^\mu
+ \alpha' N[ d \rho^2 + e^{ 2 g(\rho)}
(d\theta_1^2+ \sin^2\theta_1 d\phi_1^2)+
{1 \over 4 } \sum_a (w^a - A^a)^2 ] \right],
\\
\fl G_3= ie^{\Phi}F_3 = ie^{\Phi}\alpha'N \left[  -{1\over 4} (w^1 -A^1)\wedge (w^2 - A^2) \wedge ( w^3-A^3)  + { 1 \over 4}
\sum_a F^a \wedge (w^a -A^a) \right].\nonumber
\eea
The dilaton is $\rho$-dependent and its value at the origin is a parameter $\Phi_0$.
The gauge field $A$ is given by
\be
\label{Afield}
A={1\over 2} \left[ \sigma^1 a(\rho) d \th_1
+ \sigma^2 a(\rho) \sin\th_1 d\phi_1 +
\sigma^3 \cos\th_1 d \phi_1 \right],\ee
while the one-forms $w^a$ are
\bea
 { i \over 2} w^a \sigma^a &  = &dg g^{-1}, \qquad  \qquad\qquad  \qquad  \qquad \quad g = e^{ i \psi \sigma^3 \over 2 } e^{ i \th_2 \sigma^1 \over 2 }
e^{ i \phi_2 \sigma^3 \over 2}  ,          \nonumber \\
w^1 + i w^2 & = &e^{ - i \psi } ( d \th_2 + i \sin \th_2 d \phi_2)  ~,~~~~~~~~~~
w^3 = d \psi +\cos \th_2  d\phi_2.
\eea
The functions $g(\rho),\, \Phi(\rho),\, a(\rho)\,$ are known in analytic form only in the supersymmetric case, but for our purposes their IR behavior, i.e. their $\rho\ra0$ limit \cite{gubsermn}
\begin{eqnarray}
\fl a(\rho)\approx 1 - b \rho^2, \quad e^g(\rho)\approx\rho - ( {b^2\over 4} +{1 \over 9}) \rho^3,\quad \Phi(\rho)\approx\Phi_0 + ({b^2\over 4} + {1\over 3}) \rho^2
\end{eqnarray}
is enough.
The basic property of this model is that it is completely regular, allowing for an analysis in the deep IR region.
In this regime it is dual to the softly broken ${\cal N}=1$ SYM with $SU(N)$ gauge group, coupled to a tower of massive KK modes in the adjoint representation.
The supersymmetry is broken by a mass term for the gaugino, which is described in the supergravity solution above by the parameter $b\in (0,2/3]$; the value $b=2/3$ corresponds to the supersymmetric solution \cite{mn}, which will be included in all the following considerations.

\subsection{The Penrose limit}

The Penrose limit consists in a boost of the theory along a null geodesic of the space-time, together with a rescaling of the transverse coordinates which allows to obtain a non singular result.
Its striking property is that if the original background is a supergravity solution, also the limiting one is a solution.
As the limit generically simplifies the background, the hope is to be able to study the string theory on the resulting space-time.

The null geodesic we will choose is fixed a $\rho\sim 0$, since this is the region where the background is dual to the four dimensional field theory, and because the large charge hadrons we will study have wave function peaked at this value of the radius.
The geometry of the solution (\ref{MNsol}) is such that at $\rho = 0$ there is a vanishing two sphere and a three-sphere with finite volume.
Our geodesic will thus be taken to lie along the equator of the three-sphere defined by $\theta_2=0$, $\phi_2=\psi$.
Doing the limit carefully\footnote{We shift the flat coordinates as $e^{\Phi_0/2}L^{-1}x_{\mu}\ra x_{\mu}$, with L an arbitrary constant. The tension of the confining strings is $T_s = L^2/(2\pi\alpha')$, the glueball and KK masses are instead $M_{KK}^2\approx M_{gl}^2\approx L^2/(e^{\Phi_0}N\alpha')$. Then we take $L^2\approx N\ra\infty$, which sends the string tension to infinity while keeping $M_{KK}, M_{gl}$ fixed. Note that, thanks to the above rescaling, we avoid sending the dilaton to infinity.} gives
\bea\label{ppMNr}
\fl ds^2 = -2dx^+dx^- -m_0^2\,({b^2\over 4}u_1^2 + {b^2\over 4}u_2^2+v_1^2+v_2^2)(dx^+)^2  +d\vec{x}^{\,2} +
dz^{\,2} + du_1^2+du_2^2+ \nonumber \\
 \quad + dv_1^2+dv_2^2  ,\\
\fl G_3 = -2 i m_0 dx^+\wedge [dv_1\wedge dv_2 + {b\over 2} du_1\wedge du_2], \qquad \Phi=\Phi_0, \nonumber
\eea
where $m_0=L/\sqrt{e^{\Phi_0}\alpha'\, N}$ is the mass scale of the glueball and KK fields, which has been kept fixed in the limit.
This is an essential feature of the construction, which is meant to study objects of definite four dimensional Minkowski mass, i.e. hadrons.
As in the well known BMN case, the Penrose limit selects only those states with very large quantities associated to the null geodesic.
In our case these quantities are the mass and a symmetry charge associated to the three-sphere equator, which we call $J$.

\subsection{Spectrum}

Now we come to the first surprise of the model: the string theory on the background (\ref{ppMNr}) is solvable!
Previous attempts to study non conformal backgrounds in the Penrose limit had instead always ended up with theories with time dependent world-sheet masses, a feature that makes the solution quite complicate.

We will skip every detail and just give the spectrum. If we define $m=m_0 \alpha' p^+$, we have four massless fields ($x^i, z$) with frequencies $\omega_n=n$, and four massive fields ($u_1,u_2,v_1,v_2$) with frequencies
\be
\omega_n^u = \sqrt{n^2 + {b^2\over 4}\,m^2},\qquad \qquad 
\omega_n^v = \sqrt{n^2+ m^2}.
\ee
The 0-modes of $u_1,u_2$, together with the one of $z$, come from the mixing of the two-sphere and the radius direction on the original metric (\ref{MNsol}), while the ones of $v_1,v_2$ come from the three-sphere directions orthogonal to the null geodesic.
Note that the supersymmetry breaking parameter enters only in the $u_1,u_2$ frequencies.
Instead, all the frequencies of the worldsheet fermions are $b$-dependent
\bea
\omega_n^{I}&=& \sqrt{n^2 + {m^2\over 4}\Biggl(1+{b\over 2}\Biggr)^2}, \qquad I=1,2,3,4; \nonumber \\
\omega_n^{K}&=& \sqrt{n^2 + {m^2\over 4}\Biggl(1-{b\over 2}\Biggr)^2}, \qquad K=5,6,7,8 .
\eea
Finally, we can express the Hamiltonian and the light-cone momentum in terms of field theory quantities
\be
\label{MNH}
 H = -p_+ = E-m_0\,J, \qquad  P^+ = {m_0\over L^2}\,J ,
\ee
$E$ being the hadron mass, and 
\be\label{gei}
J=U(1)_l -{b \over 2} U(1)_{J_1} + U(1)_r,
\ee
where $U(1)_{J_1}$ is the two-sphere isometry and $U(1)_r,\, U(1)_l$ are the two $U(1)$ isometries of the three-sphere $SO(4)$.

\subsection{Vacuum energy}
The background (\ref{ppMNr}) preserves sixteen supersymmetries \cite{pandostras}. 
These are the ubiquitous, ``kinematical'' ones and give no linearly realized world-sheet supersymmetry \cite{cve}.
As a consequence of this fact, the bosonic and fermionic frequencies do not cancel and there is a non trivial vacuum energy.
It reads
\be \label{zeromn}
E_0(m)={m_0\over 2m}\sum_{n=-\infty}^{\infty}\left[ 4n + 2\omega_n^u + 2\omega_n^v - 4\omega_n^1 -4\omega_n^5 \right]
\ee
and it is always {\it negative}, also in the $b=2/3$ case!\footnote{This can be explicitly checked by first regularizing the expression above (for example through an Epstein function regularization) and then calculating it. We stress that being our model supersymmetric, there is no need \cite{noistab} to renormalize (\'a la Casimir) the bosonic and fermionic contributions in $E_0$: as always happens for supersymmetric models, all the infinities are just naturally canceled by the combination of both the contributions.} 

This fact raises naturally the problem of the stability of the theory.
On the one hand, there are preserved supersymmetries, so that one would think that the background is surely stable.
But the ordinary arguments for this to be true rely on flat space (or $(A)dS$) constructions, while our metric is not asymptotically flat (or $(A)dS$).
For the very same reason, we can't conclude that we have an instability because of the negativity of the energy.
While a complete discussion of the stability of the theory at the quantum level is a very hard problem, we can perform a classical analysis in the supergravity limit, i.e. at small $m$, where $E_0$ is given by the contribution of the zero frequencies $E_0\ra (m_0/2)[|b| +2 -2|1-(b/2)| -2|1+(b/2)|]$.
The corresponding supergravity mode, a combination of the trace of the graviton with the three-form, is stable if it has no diverging exponential behavior with time. 
It can be shown that the ground state of the above string theory, i.e. the supergravity field with light cone energy $E_0$ as above, has a dispersion relation which reads $\sqrt{2}\omega=\sqrt{2}p_y+ m_0(b -2)$
where $p_y$ is the transverse momentum which combines with $\omega$ in the light-cone momenta $p^{\pm}$.
We see that we get a real $\omega$ for every real $p_y$.
Being the dangerous part of the supergravity mode of the $Exp(i\omega t)$ type, we conclude that the theory is {\it classically stable} \cite{noistab}\footnote{Note that this criterion for the stability rules out the Type 0 string on the maximally supersymmetric pp-wave \cite{noi0}.}. 
\section{Field theory data}\label{data}

In this section we will identify some of the string modes of the previous sections with field theory states \cite{noi}.
As already stressed, these will be hadrons with very large mass and very large charge $J$.
Their constituents are the lightest KK fields, since these are the lightest objects with non-trivial $J$-charge (the gauge fields are uncharged).
In the (b)MN model there are in fact four KK scalars at the lowest level, transforming in the bifundamental of $SU(2)_l\times SU(2)_r$ and in the adjoint representation of the gauge group. 
Their mass scale is $m_0$ by construction, and their $U(1)_{J_1}$ charges are all zero, since they are all uncharged with respect to the spin connection of the two-sphere\footnote{Thus, in this \none dual solution, corresponding to wrapped five-branes, there are no massless scalars, as their $U(1)_{twist}=U(1)_r$ is non-zero, while in the \ntwo case there are two massless scalars, since for them the $U(1)_{twist}=U(1)_r+U(1)_l$ is zero \cite{noi2}.}.
Let's call them $A_{i},\,\, i=1,...,4$.
Thus, from (\ref{gei}) we see that their J charges are $1,0,0,-1$ respectively.
It is then natural to identify the string theory vacuum $|0,p^+\rangle$ with the hadron $(A_{1})^{J}|0\rangle$, $|0\rangle$ being the field theory vacuum. 
In fact, having $J$ $A_1$ components, the Hamiltonian (\ref{MNH}) for this state is $H=Jm_0-m_0*J*1=0$, equal to the string ground state value (ignore, for the moment, the vacuum energy contribution), if the mass of the KK scalar is exactly $m_0$.
The latter can be viewed as the string theory prediction for a quantity which is almost impossible to calculate in the field theory side, as it represents the mass of the scalar in the strong coupling regime and in the mean field of all the other scalars in the hadron.

Let's come to the excitations of the ground state.
The bosonic part of the string Hamiltonian can be written as \cite{pandostras}
\be
H=\frac{P_i^2}{2m_0J}+\frac{T_s}{2m_0J}(N_L+N_R)+\frac{T_s}{2m_0J}(H_0+H_L+H_R),
\ee
where $P_i,\, N_L,\,N_R$ are the three special momenta and number operators associated to the massless modes, $H_0,\,H_L,\,H_R$ are the sub-Hamiltonians of the other five modes, and $T_s$ is the gauge theory string tension.
Thus, from the first term we immediately read the three dimensional {\it non relativistic motion} of our hadrons in the three special directions!
A posteriori, this is expected, since we are probing very heavy hadrons at low energy (we are in the IR regime), but nevertheless the appearance of the precise kinetic energy for this configuration is astonishing.
The second, even more astonishing feature, comes from the $N_L,\,N_R$ contributions, which account for the string excitations in the three special directions.
They tell us that the hadrons have really string-like excitations, i.e. they are physically {\it stringy shaped} (hence the name ``Annulons'')!.

Unfortunately, the duality is much more complicated than in the BMN case, and little can be said about the remaining part of the string spectrum.
We control only two other zero-modes, namely $v,\bar{v}=v_1\pm i\,v_2$, which have exactly the same quantum numbers of the scalars $A_2,\,A_3$.
So, their action on the string ground state corresponds to the insertion of the corresponding $A$ fields in the string of $A_1$ above, and their Hamiltonian is $H=E-m_0J=m_0-m_0*0=m_0$, the value of the corresponding string oscillators.
These modes complete the so-called ``universal sector'', which is determined by the symmetry of the original background in the IR.
In fact, apart from the flat coordinates, for which we gave the interpretation above, the geodesic direction on the three-sphere is naturally dual to the ground state Annulon (the $A_1$ field), while the oscillations transverse to the geodesic, $v, \, \bar v$, are identified with $A_2,\,A_3$. 
The modes corresponding to the remaining part of the geometry, namely the vanishing two-sphere and the radius direction, have not been identified yet.

Finally, the ``would be supersymmetric partners'' of the scalars $A_1,\, A_2$, i.e. the fermions in the same chiral multiplet in the supersymmetric case, account for four of the string fermionic zero modes, two with mass ${m\over 2}(1+{b\over 2})$ and two with mass  ${m\over 2}(1-{b\over 2})$ \cite{noi}.

\section{Final remarks}\label{conc}

In this paper we have reviewed how we can study features of confining, (non) supersymmetric gauge theories that have a string dual via the Penrose limit technique.
This allows to describe some properties of certain large mass, large charge hadrons, called Annulons.

There are a number of interesting points to be clarified in the behavior of these hadrons.
Their most striking feature is that they are physically string-like.
This unexpected topic surely deserves more investigation, in order to clarify how the Annulons behave in details.
Since they are quite unusual object to find in field theory, one of the most obvious question is whether they are present only in the (broken) MN and KS models or not.
It has been shown in \cite{noi5} that the presence of the Annulons is really a generic feature of the confining field theories which have a critical string dual.
In fact, the known string duals of three and four dimensional \none SYM embedded in Type IIA and Type IIB theories have these states in their spectrum.

Finally, let's comment on the possible perturbative computations one would like to perform for these hadrons.
In the large $m$ limit the vacuum energy (\ref{zeromn}) become
\be
E_0\approx -m_0 m * A^2, \qquad \qquad A^2=const.
\label{Imn}
\ee
This takes larger and larger negative values as $m$ increases.
As such, it represents a huge correction to the energy-charge relation for the Annulons, in a regime where the field  theory should be effectively weakly coupled.
In fact, $m$ is proportional to the ratio of the current $J$ and the factor $Ne^{\Phi_0}$, call it $\lambda$ (it is large in the Penrose limit), i.e. $m^2$ is proportional to the inverse effective coupling constant, $\lambda'$ in the standard notation in the BMN case.
Thus, the large $m$ regime means weak $\lambda'$ coupling in the field theory, and the corrections above are of order 
$1/\sqrt{\lambda'}$, very large indeed.

This could be read as a signal of breaking down of the perturbative expansion in our model.
On the one hand, we could also  read it as a perturbative correction to the predicted value of $m_0$.
The general expression for the energy-charge relation can in fact be written, in analogy with the AdS/${\cal N}=4$ case \cite{tseyrot}, as
\be
E=J*m_0(1-{A^2\over \lambda} + {\cal O}({1\over \lambda^2})).
\ee
We see that this expansion is a kind of ``instanton like'' correction to the value of $m_0$.
Since the latter cannot be computed in the perturbative field theory, the presence of these corrections doesn't change much the situation.
So, in principle, we could just accept the new value of $m_0$ as given by this series as a stringy prediction, and go on to calculate, for example, the corrections to the mass of the Annulons with insertions of field corresponding to string oscillators, along the line of BMN.

On the other hand, the presence of these large corrections is somewhat disturbing, and one would eventually avoid them.
Moreover, they are the same kind of corrections one gets in the AdS/${\cal N}=4$ case considering the rotating string corresponding to only one charge \cite{tseyrot2}.
In that case, the corrections become subleading if one considers multi-charge generalizations of that solution \cite{tseyrot,tseyrot3}.
So, the hope is that also for the Annulons, considering multi-charge or charge-spin semiclassical solitons could cure the one loop sigma model corrections, making them subleading.
This is an interesting problem for a future investigation.

\ack
F. B. is partially supported by INFN.


\end{document}